\documentclass{elsart}
\usepackage{graphicx}
\usepackage{amssymb}
\begin{document}
\begin{frontmatter}

\title{Evaluating the RiskMetrics Methodology in Measuring Volatility and 
Value-at-Risk in Financial Markets}
\author[addr1]{Szil\'ard Pafka\thanksref{email1}}, 
\author[addr1,addr2]{Imre Kondor\thanksref{email2}} 
\address[addr1]{Department of Physics of Complex Systems, E\"otv\"os University,\\
P\'azm\'any P.\ s\'et\'any 1/a, H-1117 Budapest, Hungary} 
\address[addr2]{Market Risk Research Department, Raiffeisen Bank,\\
Akad\'emia u.\ 6, H-1054 Budapest, Hungary}
\thanks[email1]{syl@complex.elte.hu}
\thanks[email2]{ikondor@raiffeisen.hu}

\begin{abstract}
We analyze the performance of RiskMetrics, a widely used methodology for 
measuring market risk. Based on the assumption of normally distributed returns, 
the RiskMetrics model completely ignores
the presence of fat tails in the 
distribution function, which is an important feature of 
financial data. Nevertheless, it was commonly found that RiskMetrics
performs satisfactorily well, and therefore the technique has become
widely used in the 
financial industry. We find, however, that the success of RiskMetrics is 
the artifact of the choice of the risk measure. First, the 
outstanding performance
of volatility estimates is basically due to the choice of a very short 
(one-period ahead) forecasting horizon. Second, the satisfactory
performance in
obtaining Value-at-Risk by simply multiplying volatility 
with a constant factor is mainly
due to the choice of the particular significance level.
\end{abstract}

\end{frontmatter}


\section{Introduction}

Risk management is one of the top priorities in the financial industry today. 
A huge effort is being invested into
developing reliable risk measurement methods and sound risk
management techniques by academics and practitioners alike. 
J.\ P.\ Morgan was the first to develop a comprehensive
market risk management methodology based on the Value-at-Risk (VaR) concept 
\cite{jorion}.
Their product, RiskMetrics \cite{riskm}  has become extremely popular
and widespread. It has greatly contributed to the dissemination of
a number of basic statistical risk measurement methods and to the general acceptance 
of VaR as an industry standard.
Although backtests performed first by J.\ P.\ Morgan and later
by other market participants lent support to the RiskMetrics model, its basic assumptions
were shown to be questionable from several points of view
(see e.g.\ \cite{carol}). 
Moreover, the existence of fat tails in real market data
(see \cite{econoph,bouchaud} for a discussion)
is in a clear conflict with RiskMetrics' 
assumption of normally distributed returns, which
can lead to a gross underestimation of risk.
Furthermore, serious doubt has recently been raised as to the stability and information content of 
the empirical covariance matrices used by the model for calculating the risk of 
portfolios \cite{rmt1,rmt2,rmt3}.

Accurate evaluation of risks in financial markets
is crucial for the proper assessment and efficient mitigation of risk.
Therefore, it is important to see to what extent widespread methodologies
like RiskMetrics are reliable
and what their possible limitations are.
In particular, we try to 
understand why, despite the evident oversimplifications embedded in the model, 
it can perform satisfactorily. We will argue that the apparent success of RiskMetrics
is due basically to the way risk is quantified 
in this framework, which does not necessarily mean that this particular risk measure is
the most adequate one. 

The paper is organized as follows. In Section \ref{sec:overview} 
a sketchy overview of the RiskMetrics methodology will be presented. In Section 
\ref{sec:success} the reasons for the success of RiskMetrics will be discussed.
Finally, Section \ref{sec:concl} is a short summary.


\section{Overview of the RiskMetrics Methodology}
\label{sec:overview}

It is well-known that daily returns 
are uncorrelated whereas the squared returns are strongly
autocorrelated. As a consequence, periods 
of persistent high volatility are followed 
by periods of persistent low volatility,
a phenomenon known as ``volatility clustering''. These features are incorporated 
in RiskMetrics by choosing a particular autoregressive moving average process to model the price process (see below). Furthermore, RiskMetrics makes the very strong 
assumption that returns 
are conditionally\footnote{Conditional here means conditional on the
information set at time $t$, which usually consists of the past
return series available at time $t$.}
normally distributed. Since the standard 
deviation of returns is usually much higher than the mean, the latter is 
neglected in the model\footnote{The typical yearly mean return in equity 
markets is 5\%, while the typical standard deviation is 15\%. For
a time horizon of one day, however, the mean becomes 5\%/250=0.02\%, while 
the standard deviation becomes $15\%/\sqrt{250}\approx 0.95\%$, i.e.\ 
about 50 times larger than the mean.}, 
and, as a consequence, the standard deviation remains the only parameter of 
the conditional probability distribution function.
In order to avoid the usual problems related to the uniformly weighted moving averages,
Riskmetrics uses the so called exponentially weighted moving average 
(EWMA) method \cite[pp.\ 78--84]{riskm} which is meant to represent the finite
memory of the market. Accordingly, the estimator for the volatility is chosen
to be
\begin{equation}
\label{eq:vol.estim}
\sigma_{t+1|t}^2=\frac{\sum_{\tau=0}^{\infty} \lambda^{\tau}r_{t-\tau}^2}
{\sum_{\tau=0}^{\infty} \lambda^{\tau}}=
(1-\lambda)\sum_{\tau=0}^{\infty} \lambda^{\tau}r_{t-\tau}^2,
\end{equation}
where $\lambda$ is a parameter of the model ($0<\lambda<1$).
The notation $\sigma_{t+1|t}$ emphasizes that the volatility estimated on a
given day ($t$) is actually used as a predictor for the volatility of the next 
day ($t+1$). The daily VaR at confidence level $p$ (e.g.\ 95\%) can then be
calculated (under the normality assumption) by multiplying $\sigma_{t+1|t}$
with the $1-p$ quantile of the standard normal distribution. Moreover, this
technique can be used to measure the risk of individual assets and
portfolios of assets as well.
For linear portfolios (i.e.\ containing no options) the usual method to
obtain the volatility is to estimate the covariance
matrix of asset returns, element-by-element, 
using the EWMA technique and then calculate 
the portfolio volatility as $\sigma_p^2=\sum_{i,j} w_i\,w_j\,\sigma_{ij}$, 
where $w_i$ is the weight of asset $i$ in the portfolio. Equivalently,
however, one can calculate the return on the portfolio first, and then 
apply the EWMA technique directly to the whole portfolio \cite{zangari}.
Finally, the value of the parameter $\lambda$ is determined by an optimization
procedure. On a widely diversified international portfolio, RiskMetrics 
found that the value $\lambda=0.94$ produces the best backtesting results
\cite[pp.\ 97--101]{riskm}. 


\section{Why Does RiskMetrics Work?}
\label{sec:success}

In order to explain the reasonably successful performance of RiskMetrics, first
we recall the work by Nelson \cite{nelson} who showed that
even misspecified models can estimate volatility rather accurately. 
More explicitly, in \cite{nelson} it is shown that if the return generating 
process is well approximated by a diffusion, a broad class of even
misspecified ARCH models\footnote{
See \cite{arch} for a survey on ARCH type models 
(e.g.\ ARCH, GARCH, IGARCH, etc.).} 
can provide consistent estimates of the conditional volatility. 
Since RiskMetrics can be considered as an IGARCH(1,1) 
model\footnote{
In the IGARCH(1,1) model, returns are generated by the
following stochastic process: $r_t=\sigma_t \varepsilon_t$, where
$\varepsilon_t\sim$ i.i.d.(0,1), $\sigma_t^2=\omega+\beta_1\,\sigma_{t-1}^2+
\alpha_1\,\varepsilon_{t-1}^2$ and $\alpha_1+\beta_1=1$.
Since from Eq.\ (\ref{eq:vol.estim})
$\sigma_{t+1|t}^2=\left(1-\lambda\right)\,\left(r_t^2+\lambda\,
\frac{\sigma_{t|t-1}^2}{1-\lambda}\right)$, it follows that
$\sigma_{t+1|t}^2=\lambda \sigma_{t|t-1}^2+(1-\lambda) r_t^2$
just as in the IGARCH(1,1) model ($\omega=0$, $\beta_1=\lambda$, 
$\alpha_1=1-\lambda$).}
the results of \cite{nelson} offer a natural explanation for 
the success of RiskMetrics in estimating volatility. 
Actually, in the RiskMetrics framework, this 
estimate is used as a one-day ahead volatility forecast, nevertheless it 
seems that this does not significantly worsen its accuracy. However,
if one uses this estimate to calculate (as often required by regulators)
a multiperiod forecast using the simple ``square-root-of-time'' 
rule\footnote{
The rule says that
$\sigma_{t+h|t}=\sqrt{h}\cdot\sigma_{t+1|t}$, which is derived on the 
basis of the assumption of uncorrelated returns \cite[pp. 84--88]{riskm}.},
the quality of the forecast is bound to decline with the number of
periods. 
A comparative study of the rate of deterioration of these forecasts with
time within the RiskMetrics model resp.\ other, more sophisticated volatility
models is an interesting topic for further research.

\begin{figure}
\begin{center}
\includegraphics[scale=0.48,angle=-90]{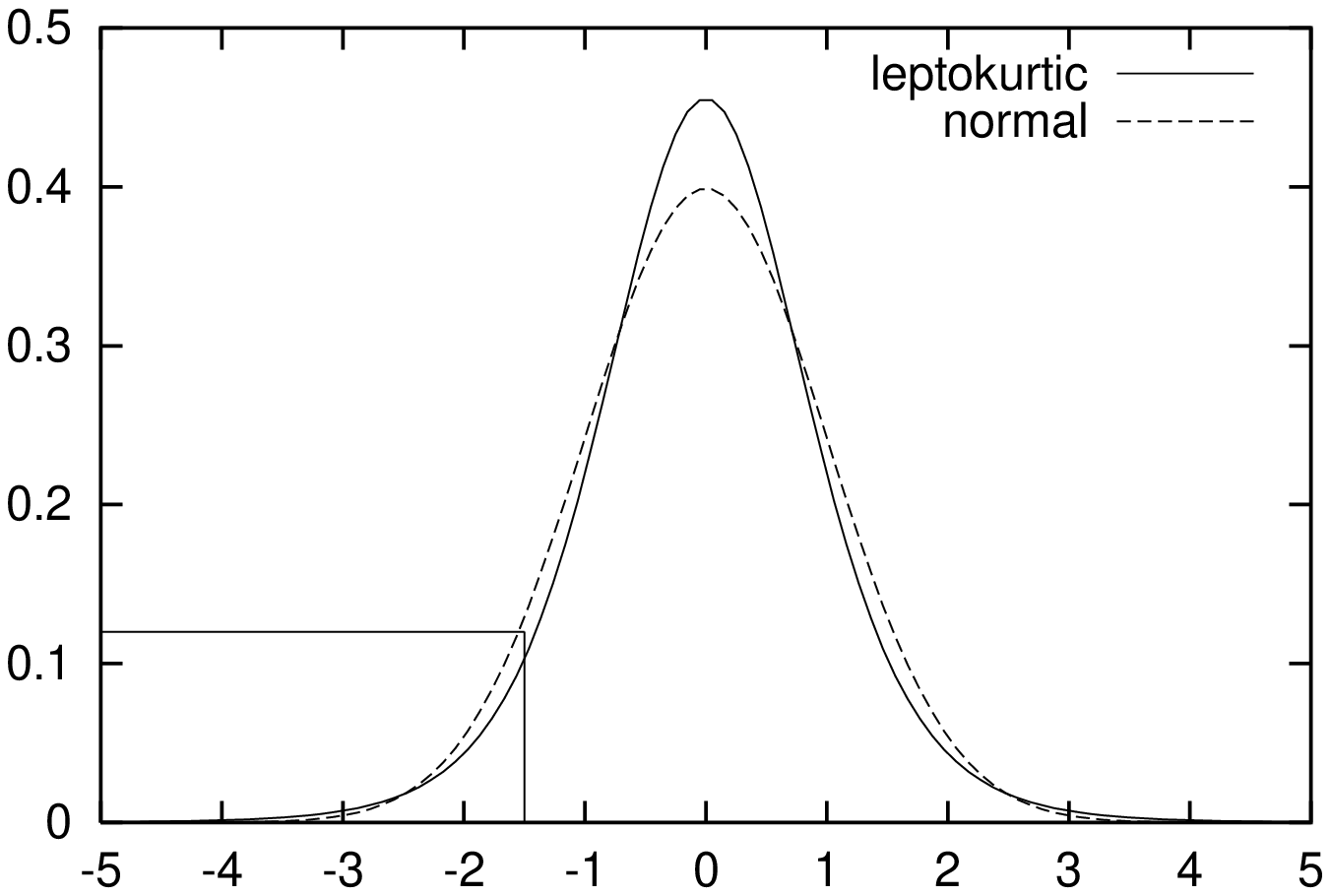}
\includegraphics[scale=0.48,angle=-90]{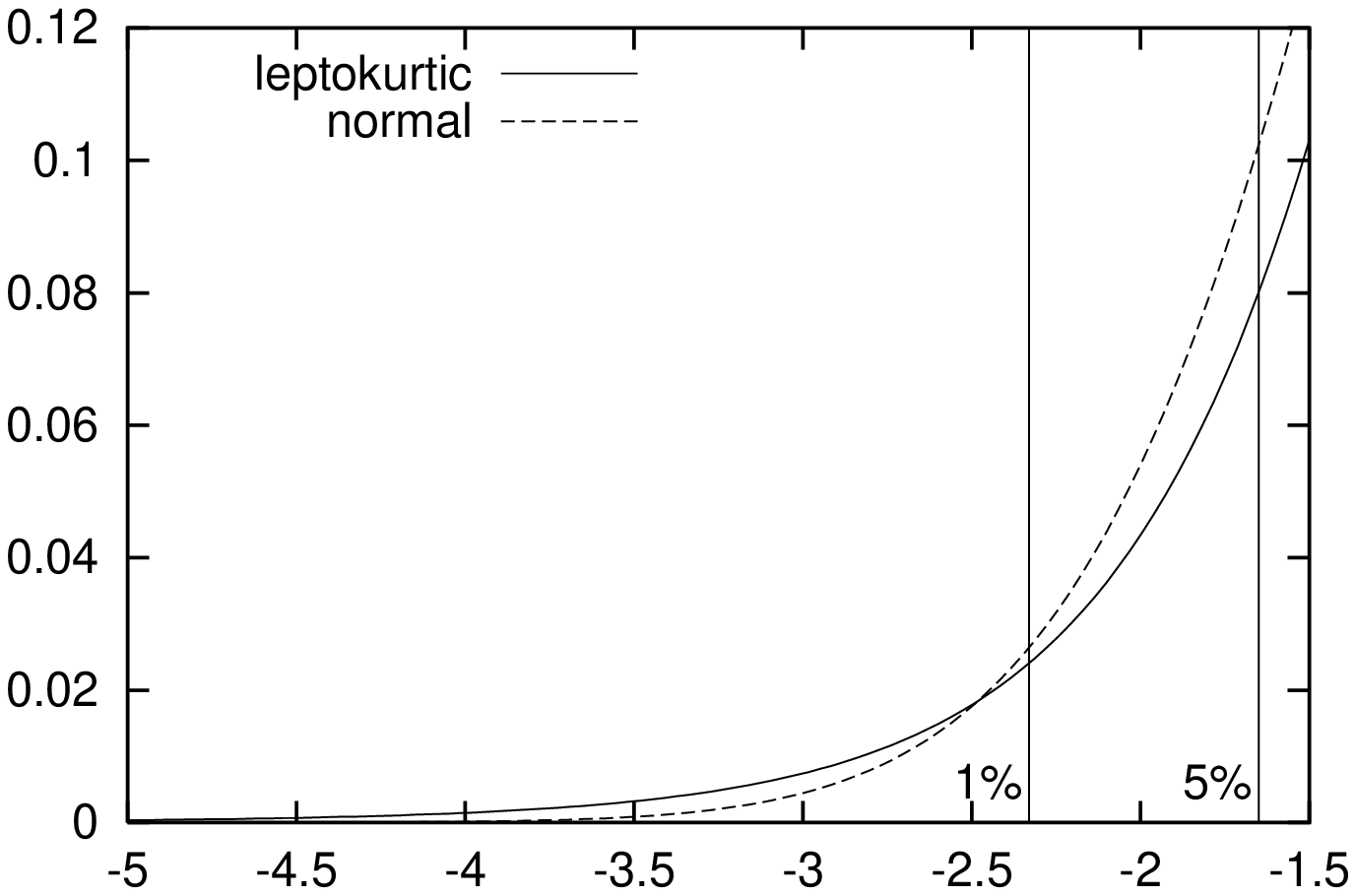}
\end{center}
\caption{Probability distribution function (PDF) of the
leptokurtic Student $t$
distribution with 7 degrees of freedom (scaled to unit variance)
and the standard normal distribution along with vertical lines at 
-1.65 and -2.33 corresponding to the 5 and 1\% quantiles of the standard
normal distribution.
\label{fig:fig1}}
\end{figure}

Let us turn to the apparent success of RiskMetrics in estimating 
the VaR now. The model sets the confidence level at 95\%. 
The prescription to obtain this 5\% quantile is to simply
multiply the volatility estimate by 1.65 (as if returns were conditionally
normally distributed). Why such a recipe can pass the backtests
can be understood if one analyzes numerically
the 5\% quantiles of a few leptokurtic (fat tailed) distributions. 
It is very often found that despite the presence of fat tails, for many 
distributions the 5\% quantile is roughly -1.65 times the standard 
deviation.   
For example, the 5\% quantile of the Student $t$ distribution with 7
degrees of freedom (which is leptokurtic and has a kurtosis of 5 
similar to the typical kurtosis of returns in financial markets) 
is -1.60, very close to -1.65, or, conversely, the -1.65 percentile is 
4.6\%. This is illustrated in Fig.\ \ref{fig:fig1}, where the 
PDF's of the Student and the normal distributions 
are shown. It can be seen from the figure that the areas
under the two PDF's to the left of the -1.65 line are roughly equal.
We also analyzed the empirical frequencies of RiskMetrics 95\% VaR violations 
which correspond to returns $r_{t+1}<-1.65\,\sigma_{t+1|t}$, where
$\sigma_{t+1|t}$ is the RiskMetrics volatility estimate obtained on day $t$. 
For the 30 stocks of the Dow Jones Industrial Average (DJIA), which are
among the most liquid stocks traded on the New York Stock Exchange, it was
found\footnote{
The sample consisted of the daily closing prices of the 30 stocks of the
DJIA from August 1996 to August 2000, about 1000 data points for each stock.}
that the VaR violations frequency
was $4.7\pm0.6\%$, i.e.\ close to 5\%. 
This explains 
why RiskMetrics is usually found so successful in evaluating risk 
(which it {\it defines} as the VaR at 95\% confidence).

It is evident, however, that for higher 
significance levels (e.g.\ 99\%) the effect of
fat tails becomes much stronger, and therefore 
the VaR will be seriously underestimated if one assumes normality.
For example, the 1\% quantile of the Student $t$ distribution considered
above is -2.54, significantly larger than under the normality assumption
(-2.33), while the percentile corresponding to -2.33 is 1.43\%. This can
also be seen from Fig.\ \ref{fig:fig1}. Furthermore, for the DJIA data
considered above, the rejection frequencies 
were $1.4\pm0.3\%$, significantly
larger than 1\%.
These effects are even more pronounced for a truncated L\'evy distribution
(TLD). For example, a Koponen-like \cite{koponen} TLD with L\'evy exponent
$\nu=1.50$, scale parameter $c=0.50$ and truncation parameter
$\lambda=0.17$, which provides an excellent fit to the Budapest Stock
Exchange Index (BUX) data, has a 5\% quantile equal to $1.76\sigma$
whereas the 1\% quantile is already $3.75\sigma$! \cite{balazs}

Therefore, it can be concluded that the satisfactory
performance of RiskMetrics in estimating
VaR is mainly the artifact of the choice of the significance level of 95\%.
However, existing capital adequacy regulations require 99\% confidence,
and at this level RiskMetrics systematically underestimates risk.


\section{Summary}
\label{sec:concl}

In this paper we analyzed the performance of RiskMetrics, perhaps
the most widely used methodology for measuring market risk. 
The Riskmetrics model is based on the unrealistic assumption of normally 
distributed returns, and completely ignores the presence of fat tails in the 
probability distribution, a most important feature
of financial data. For this reason, one would expect the model to 
seriously underestimate risk. However, it was commonly found by market
participants that RiskMetrics performed satisfactorily well and 
this helped the method to quickly become a standard in risk measurement. 
Nevertheless, we found that the success of RiskMetrics is actually
the artifact of the choice of the risk measure: the effect
of fat tails is minor when one calculates Value-at-Risk at 95\%, however,
for higher significance levels fat tails in the distribution of returns
will make the simple RiskMetrics rule of calculating VaR to underestimate risk.

RiskMetrics has played and continues to play an extremely useful role in
disseminating risk management ideas and techniques, even if oversimplified.
It is available free of charge and, coming with a careful documentation,
it is completely transparent and amenable for a study like the present one:
its limitations can be explored and, given sufficient resources, overcome.
This is far from being the case with the overwhelming majority of the 
commercially available risk management systems which incorporate 
at least as strong simplifications as RiskMetrics, but coming in the
the form of ``black boxes,'' are completely impossible to modify.
The continuing dominance of the Gaussian paradigm in risk management
software packages represents an industry-wide model risk.

\section*{Acknowledgements}

It is a pleasure to thank B.\ Janecsk\'o for useful interactions.
This work has been supported by the Hungarian National Science
Found OTKA Grant No.\ T 034835.



\begin{thebibliography}{99}

\bibitem{jorion}
Jorion P.\ (1997). {\it Value at Risk: The New Benchmark for
Controlling Market Risk}, Chicago: Irwin

\bibitem{riskm}
RiskMetrics Group (1996). {\it RiskMetrics -- Technical Document},
New York: J.P.\ Morgan/Reuters

\bibitem{carol}
Alexander C.\ (1996). ``Evaluating the Use of RiskMetrics as a 
Risk Measurement Tool for Your Operation: What Are Its Advantages
and Limitations?'' {\it Derivatives: Use Trading and Regulation},
{\bf 2}, 277--285

\bibitem{econoph}
Mantegna R.\ N., H.\ E.\ Stanley (1999). {\it An Introduction to 
Econophysics: Correlations and Complexity in Finance},
Cambridge: Cambridge UP

\bibitem{bouchaud}
Bouchaud J.-P., M.\ Potters (2000). {\it The Theory of Financial Risk:
From Statistical Physics to Risk Management}, Cambridge: Cambridge UP

\bibitem{rmt1}
Laloux L., P.\ Cizeau, J.-P.\ Bouchaud, M.\ Potters (1999). ``Noise
Dressing of Financial Correlation Matrices,'' {\it Physical Review
Letters}, {\bf 83}, 1467--1471

\bibitem{rmt2}
Plerou V., P.\ Gopikrishnan, B.\ Rosenow, L.\ A.\ N.\ Amaral, H.\ E.\
Stanley (1999). ``Universal and Nonuniversal Properties of Cross
Correlations in Financial Time Series,'' {\it Physical Review Letters},
{\bf 83}, 1471--1474

\bibitem{rmt3}
Galluccio S., J.-P.\ Bouchaud, M.\ Potters (1998). ``Rational Decisions,
Random Matrices and Spin Glasses,'' {\it Physica A}, {\bf 259}, 449--458

\bibitem{zangari}
Zangari P.\ (1997). ``A General Approach to Calculating VaR without
Volatilities and Correlations,'' {\it RiskMetrics Monitor}, 
2nd Quarter, 19--23

\bibitem{nelson}
Nelson D.\ (1992) ``Filtering and Forecasting with Misspecified
ARCH Models: Getting the Right Variance with the Wrong Model,''
{\it Journal of Econometrics}, {\bf 52}, 61--90

\bibitem{arch}
Bollerslev T., R.\ F.\ Engle, D.\ B.\ Nelson (1994). ``ARCH models,''
in Engle R.\ F., D.\ L.\ McFadden (eds.), {\it Handbook of
Econometrics IV}, New York: North-Holland, 2959--3038

\bibitem{koponen}
Koponen I.\ (1995). ``Analytic Approach to the Problem of Convergence
of Truncated L\'evy Flights towards the Gaussian Stochastic Process,''
{\it Physical Review E}, {\bf 52}, 1197--1199

\bibitem{balazs}
Janecsk\'o B.\ private communication


\end{thebibliography}
\end{document}